# Development and evaluation of a system to express a sense of telekinesis in VR

Shingo Nakaya[1], Yudai Hirota[1], Sho Sakurai[1], Takuya Nojima[1], Koichi Hirota[1]

[1] The University of Electro-Communications, Tokyo, Japan

({shingo.nakaya, hirota_yudai, sho, hirota}@vogue.is.uec.ac.jp, tnojima@nojilab.org)

**Abstract** --- Telekinesis is the ability to manipulate remote objects without direct physical contact. In fictional works, telekinesis users are often depicted as controlling objects using their hands and other body parts as if by will alone. Such depictions suggest that users experience a sense of agency over the object despite not physically touching it. In this study, we developed a VR method to simulate telekinesis and investigated whether it is possible to achieve a sense of physical sensation and agency comparable to that portrayed in fiction.

## 1 INTRODUCTION

In fiction, telekinesis—the ability to control distant objects without physical contact—is often depicted. Operators control objects by hand movements, despite no direct contact, and also experience sensations. Although this is impossible in reality, VR studies have simulated similar manipulation. For example, hand movements can be remapped onto an avatar to reach objects [1], or miniature versions of objects are manipulated to control them remotely [2]. Another study [3] mimicked telekinesis, using head movements to control objects after selection with a controller. However, participants still touched the object or interacted with it through some medium, unlike in fictional depictions where the hand is completely isolated.

When we manipulate an object, the perception of changes in both the object and the body, a sense of agency (SA) [4] arises, in which "it is not anyone else but myself who is causing a particular motion. This arises when the expected and actual movements of the object align [5]. But does SA arise when the hand and object are completely isolated, as in telekinesis? If so, how does it occur?

This research explores these questions and aims to create a sensation of operating untouched objects akin to direct contact, referred to as the "sense of telekinesis" (ST). By analyzing telekinesis in fiction, the authors developed a VR system to replicate key elements necessary for ST. The goal is to verify if the system effectively induces ST and to investigate its effects on both ST and SA, ultimately establishing a new method for simulating telekinesis and offering a unique interaction experience.

## 2 A SYSTEM EXPRESSES A SENSE OF TELEKINESIS

### 2.1 Analysis of the description of telekinesis

To develop a system that expresses a ST, the authors surveyed 72 fictional works on telekinesis, analyzing object manipulation methods and the sensations described.

When manipulating an object in an arbitrary direction, the operator selected an object by looking at it (53 works), opened their hands (58 works), extended their arms to the object, and moved their hands (64 works). In these depictions, the movement of the hand and the object are synchronized. The operator's sensations are described as involving strain and concentration, with details like a trembling hand from tension (41 works) and intense, unblinking focus on the object (45 works). Furthermore, some works depicted the aura of energy interfering with the object (40 works).

Based on these findings, we considered that the following three processes are important in generating a ST: (1) "Concentrate" one's attention on a target by gazing on it for generating energy, (2) experiencing "strain" in the hands to convey energy toward the manipulated object, and (3) perception of a "sense of energy (SE)" as it gathers in the target, arms, and hands.

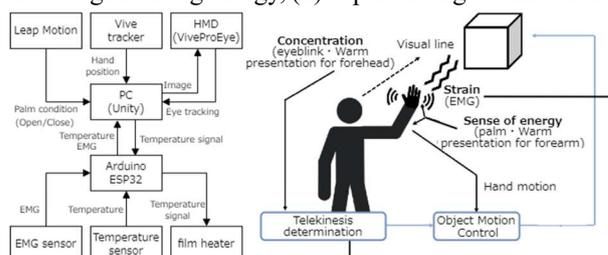

Fig.1 System configuration and images of the experience

## 2.2 System Implementation

Figure 1 illustrates the configuration and experimental depiction of a system that embodies the factors of the ST described in the previous section.

When the operator focuses their gaze and palm on the target while straining their hand and concentrating, they are considered to be in a telekinetic state. Currently, visual effects are applied to both the object and the avatar's arms, and thermal stimuli are presented on the forearms, forehead, and palms to convey a SE. As the operator moves their hand in this state, the object follows the movement.

### 2.2.1. Object Manipulation

Hand position is tracked by ViveTracker (HTC) attached to the palm of the hand, and hand opening/closing is tracked by LeapMotion (Ultraleap) attached to the front of the HMD. When the operator opens and moves his hand toward the object, the object follows the movement of the hand. The displacement of the object per frame is determined by the operator's sensitivity factor $k$, the object's speed $s$ is determined by the distance between the hand and the object, and the object's direction of movement $dir(t)$, as shown in equation (1) below.

$$\Delta x = k s\, dir(t) \quad (1)$$

$s$ is determined by the amount of hand movement $m(t)$ and the distance between the hand and the object $dist_{hx}$, as shown in equation (2).

$$s = m(t) dist_{hx} \quad (2)$$

Next, Equation (3) shows that the amount of hand movement at times $t$ and $t$-1 is compared, and the one with the greater amount is used to reflect the operator's intention.

$$m(t) = \begin{cases} m(t-1) & (m(t-1) > m(t) \text{ and } sim) \\ m(t) & (\text{otherwise}) \end{cases} \quad (3)$$

The inner product is used to determine the similarity of the direction of movement, as shown in Equation (4).

$$sim = \begin{cases} True & \boldsymbol{dir}(t) \cdot \boldsymbol{dir}(t-1) > sim_{th} \\ False & \boldsymbol{dir}(t) \cdot \boldsymbol{dir}(t-1) < sim_{th} \end{cases} \quad (4)$$

The direction of the object's movement is determined as shown in Equation (5).

$$\boldsymbol{dir}(t) = \begin{cases} D|(\boldsymbol{x}(t-1) - \boldsymbol{h}(t)| & (m_d > m_{vh} \text{ and } m_d > m_{th}) \\ \boldsymbol{h}(t) - \boldsymbol{h}(t-1) & (m_{vh} > m_d \text{ and } m_{vh} > m_{th}) \\ \boldsymbol{dir}(t-1) & (\text{otherwise}) \end{cases} \quad (5)$$

where $m_d$ and $m_{vh}$ represent the amount of hand movement in the forward/backward and up/down/left/right directions, respectively. $m_{th}$ is the threshold to determine whether the hand is moving, and $D$ indicates hand moving is forward/backward.

### 2.2.2 Measurement of concentration state

Based on the study showing that blink frequency decreases during concentration [6] and the previous depiction analysis, concentration was measured using blink intervals via ViveProEye (HTC). Since blinks decrease by 40% during concentration [7], the concentration threshold, $C_{th}$, is set to 1.67 times the average blink interval measured over 60 seconds in the user's normal state.

### 2.2.3 Strain Measurement

The strain was determined by attaching an EMG sensor (MyoScan, ALTs) to the arm and measuring the EMG signals. The sensor was connected to an M5Stick, which converted the EMG waveform to EMG strength and transmitted it to Unity via Bluetooth. The sampling rate was set to 2 kHz. During the waveform-to-intensity conversion process, the DC component was first removed from the waveform. Next, the absolute values were calculated. The EMG signal strength is normalized using Equation (6).

$$F'_t = \frac{F_t - F_{min}}{F_{max} - F_{\min}} \quad (6)$$

where $F_t$ is the strength of the EMG, $F_{max}$ and $F_{min}$ are the maximum and minimum values of $F_t$, and $F'_t$ is the normalized value of $F_t$ scaled to the range of 0 to 1.

### 2.2.4 Presentation of thermal stimuli

Wallfront film heaters were used on the forearms and foreheads, whereas fence-up film heaters were used on the palms. NTC thermistors (Murata Manufacturing) (accuracy±0.5 °C) were used as temperature sensors. The film heaters were controlled by temperature sensors to maintain the skin surface temperature at a normal level plus 2 °C, based on the temperature threshold at which the skin starts to feel warm [8]. However, for safety reasons, the temperature was controlled to ensure it did not exceed 40 °C.

## 3 EVALUATION EXPERIMENT

### 3.1 Purpose of the Experiment

In this experiment, we examined the effects of the "concentration", "strain", and "SE" on the ST, as described in the previous section.

## 3.2 Experimental Design and Procedures

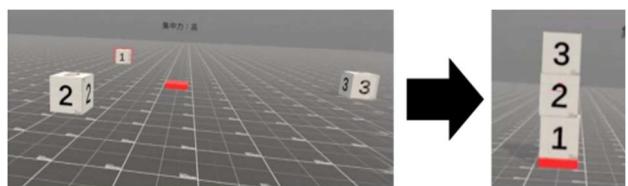

Fig.2 Object movement task

Using the system environment described in Section 2.2, we designed a task in which the participants use telekinesis to move blocks within the VR space. As shown in Figure 2, three blocks are placed, and participants are instructed to stack them in a specified order at a target position. In this task, the conditions for during telekinesis use were set as follows: "Concentration" (Yes/No), "Strain" (Yes/No), "SE (Yes/No), resulting in a total of 8 possible conditions. The device's behavior under each condition was assessed as follows: For the "concentration" condition, the device judges the participant to be in the concentration state when the average of the most recent five blink intervals with the participant's gaze on the object exceeded the threshold value $C\_th$. For the "no concentration" condition, the judgment was based solely on whether the gaze was directed at the object. For the "strain" condition, the device judges the participant to be in the strain state when the normalized values of the EMG potentials $F\_t^{\wedge\prime}$ was greater than the threshold value $F\_th$ set for each experiment. In the "no strain" condition, the EMG potentials were not used. In the "SE" condition, thermal stimuli were presented during object manipulation, while thermal stimuli were not used in the "no SE" condition.

Before experiment, participants completed a questionnaire on their dominant hand, HMD use, telekinesis concepts, and related media knowledge. They then wore the HMD, EMG sensor, film heater. After practicing, they performed the main task 5 times per condition. Post-task, they completed a questionnaire (Table 1), rating SA on a 7-point Likert scale and ST using the VAS method. In each of the eight conditions, participants completed tasks and questionnaires. Each condition was tested once in randomized order to counteract order effects. Ten participants (9 males, 1 female) took part in the experiment.

## 3.3 Experimental results

Due to space limitations, we focus on evaluating SA and ST. A Shapiro-Wilk test revealed a lack of normality in some questionnaire scores, so an aligned rank transformation (ART) was applied before conducting a three-factor ANOVA. Figure 3 presents the results where a main effect was observed. For SA: A1: SE ($F(1, 63) = 0.583$, $p = 0.448$), A3: concentration ($F(1, 63) = 4.88$, $p = 0.031$), strain ($F(1, 63) = 4.88$, $p = 0.031$), A4: concentration ($F(1, 63) = 8.29$, $p = 0.005$), A6: strain ($F(1, 63) = 5.78$, $p = 0.019$). All interactions were non-significant.

For ST: T1: concentration ($F(1, 63) = 164.3$, $p < 0.001$), T2: strain ($F(1, 63) = 195.1$, $p < 0.001$), T3: concentration ($F(1, 63) = 10.9$, $p = 0.002$), strain ($F(1, 63) = 5.29$, $p = 0.024$), SE ($F(1, 63) = 5.36$, $p = 0.024$), T5: strain ($F(1, 63) = 4.01$, $p = 0.049$), SE ($F(1, 63) = 8.11$, $p = 0.006$), T6: strain ($F(1, 63) = 10.7$, $p < 0.001$), T7: SE ($F(1, 63) = 5.46$, $p = 0.023$). Only T2 showed a significant second-order interaction (concentration × strain × SE: $F(1, 63) = 4.76$, $p = 0.033$).

## 4 DISCUSSION

The main effects for the "concentration" factor in T1, the "strain" factor in T2, and all three factors in T3 were observed in the evaluation of the ST. The evaluation scores were significantly higher in the Yes condition compared to the No condition for all factors. These results indicate that the design of the system effectively represented "concentration," "strain," and "SE," respectively. Additionally, a second-order interaction of T2 suggests "concentration" and "SE" may enhance the persuasiveness of the scenario where the subject moves the object through strain. The main effects of "strain" were significant for T4 and T6, "strain" and "SE" for T5, and "SE" for T7. In each case, the evaluation values were significantly higher in the yes condition than in the no condition. The results from T4 and T6 indicated that the design involving strain effectively produced the sensation of manipulating objects through telekinesis and applying force to an object without physical contact. T7 suggested that the thermal stimulus enhanced the sensation of touching a remote object. These findings indicate that both strain and SE are valuable for expressing ST. Additionally, the results of T5 show that, while strain and SE can individually convey a sense of telekinesis, their combination enhances the quality of this sensation. Therefore, incorporating both the strain and SE into the procedure is crucial for generating a convincing ST.

Table 1 Questionnaire during the experiment

| Sense of Agency | |
|---|---|
| A1 | I felt as if the object moved as I wished, as if it followed my will. |
| A2 | I expected the object to respond my hand movement in the same way. |
| A3 | Operation was (1 difficult, 7 easy). |
| A4 | I felt I could manipulate the object as I wished. |
| A5 | I felt I was manipulating the object as if it were part of my own body. |
| A6 | I felt as if I was operating the object with my own hands. |
| A7 | I felt as if my own hands were linked to the object. |
| Sense of Telekinesis | |
| T1 | I needed concentration in my operations. |
| T2 | I needed to strain in my operations. |
| T3 | I needed to sense of energy in my operations. |
| T4 | I felt a sense of reality (those things) as manipulation by telekinesis. |
| T5 | The sensations during the operation were like I was invoking telekinesis. |
| T6 | I felt a sense of force applied to the object. |
| T7 | I felt a sense of connection between my hand and the object. |

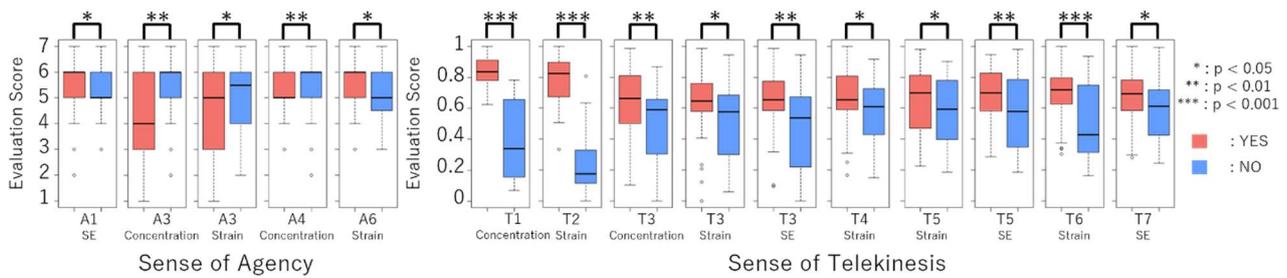

Fig.3 Evaluation values for each factor with a main effect found in the post-task questionare

In the evaluation of SA, main effects were observed for the "SE" in A1 and "strain" in A6. In both cases, the evaluation values were significantly higher for the yes condition than for the no condition. This can be considered to have enabled the user to feel as though they were manipulating the object with their intention, even without directly touching it, through a system design where concentrating force into the operator's hand to move the object results in a sensation of energy.

On the other hand, significant main effects were observed for "concentration" and "strain" in A3, and for "concentration" in A4. Evaluation values were significantly higher in the No condition compared to the Yes condition for both factors. This may be due to the system not considering the object's physical characteristics. For example, manipulating a heavy object in real life requires concentration and strain. However, the system's design may not have produced a sensation of weight that required these factors. We believe this discrepancy increased the difficulty of operation and inhibited the generation of a sense of telekinesis (ST).

## 5 Conclusion

In this study, we developed a VR system to represent the sense of telekinesis (ST) using three factors: "concentration", "strain", and "sense of energy (SE)", based on the depictions from fictional works. We evaluated the effects of these factors on ST and sense of agency (SA) towards remote objects. The results indicate that "strain" and "SE" significantly contribute to the generation of ST and SA. On the other hand, "concentration" may contribute to the generation of ST by taking into account the load related to real object manipulation. Future studies will examine the relationship between the physical characteristics of the object and these three factors to investigate new ways of inducing ST to enhance its realism.